\documentclass[12pt,a4paper]{article}
\pdfoutput=1

\usepackage[utf8]{inputenc}
\usepackage[T1]{fontenc}
\usepackage[margin=2.0cm]{geometry}
\usepackage[pdfstartview={FitH},bookmarks=false,linktoc=page,colorlinks=true,linkcolor=blue,citecolor=blue,urlcolor=blue]{hyperref}
\usepackage[numbered]{bookmark}
\usepackage{mathtools}
\usepackage{amssymb}
\usepackage{graphicx}
\usepackage[font=small,labelfont=bf]{caption}
\usepackage{authblk}
\usepackage{cite}
\usepackage{dsfont}
\usepackage{xcolor}
\usepackage[titles]{tocloft}
\usepackage{float}
\usepackage{subcaption}
\usepackage{color}
\usepackage{tikz}
\usepackage{ifthen}
\usepackage{amsmath}
\usepackage[warn]{textcomp}
\numberwithin{equation}{section}
\allowdisplaybreaks
\usepackage{multirow}

\usepackage[displaymath, mathlines]{lineno} 

\begin{document}

\title{\vspace{2cm}\textbf{Thermodynamic Stability of the Stationary Lifshitz Black Hole of New Massive Gravity}\vspace{1cm}}

\author[a,1]{K. Kolev}
\author[a,2]{K. Staykov}
\author[a,b,2]{T. Vetsov}

\affil[a]{\textit{Department of Physics, Sofia University,}\authorcr\textit{5 J. Bourchier Blvd., 1164 Sofia, Bulgaria}
	
	\vspace{-10pt}\texttt{}\vspace{0.0cm}}

\affil[b]{\textit{The Bogoliubov Laboratory of Theoretical Physics, JINR,}\authorcr\textit{141980 Dubna,
		Moscow region, Russia}
\vspace{10pt}\texttt{}\vspace{0.1cm}}

\affil[1]{\texttt{kolev.kalin@gmail.com}}
\affil[2]{\texttt{kstaykov,vetsov@phys.uni-sofia.bg}}

\date{}
\maketitle


\begin{abstract}
In this paper we investigate the thermodynamic properties of the stationary Lifshitz black hole solution of New Massive Gravity. We study the thermodynamic stability from local and global  point of view. We also consider the space of equilibrium states for the solution within the framework of Thermodynamic Information Geometry.  By investigating the proper thermodynamic metrics and their curvature invariants we find a set of restrictions on the parameter space and the critical points indicating phase transitions of the system. We confirm our findings by analytical analysis of the geodesics on the space of equilibrium states.
\end{abstract}

\vspace{0.5cm}
\vspace{0.5cm}
\thispagestyle{empty}

\noindent\rule{\linewidth}{0.75pt}
\vspace{-0.8cm}\tableofcontents
\noindent\rule{\linewidth}{0.75pt}

\section{Introduction}\label{sec: Introduction}

In the recent years alternative gravitational theories in three dimensions with black hole solutions have become very attractive area of research. This is mainly due to the  impressive gauge/gravity correspondence, where such solutions are dual to two-dimensional quantum field theories at finite temperatures. In general, there are two ways to construct three-dimensional models of gravity. 

In the first approach, one add topological Chern-Simons terms to the standard Einstein-Hilbert action. The resulting theory is known as Topologically Massive Gravity (TMG) proposed by Deser, Jackiw and Tempelton in \cite{PhysRevLett.48.975,DESER1982372}. Here, the propagating degree of freedom is a massive graviton. Among other exact solutions the theory also admits the famous Banados-Teitelboim-Zanelli (BTZ) black hole \cite{Banados:1992wn}. Since its proposal, TMG has been extensively studied in the literature and  different features of the model has already been uncovered (see for example \cite{Li:2008dq, Birmingham:2010gj, Sachs:2008gt, Park:2008yy, Blagojevic:2008bn, Solodukhin:2005ah, Anninos:2008fx, Dimov:2019fxp}). An extension of TMG with additional curvature squared term in the field equation has been proposed in  \cite{Bergshoeff:2014pca} called Minimal Massive Gravity (MMG). For certain range of the parameters MMG admits positive energy of the bulk graviton and positive central charges of the dual conformal field theory (CFT), thus avoiding the bulk-boundary unitarity problem arising in TMG. 

In the second approach, the Einstein-Hilbert action is modified by higher-derivative correction terms, which give rise to the three-dimensional New Massive Gravity theory (NMG) \cite{PhysRevLett.102.201301}. In contrast to TMG, NMG is a parity preserving theory \cite{PhysRevLett.102.201301} and with certain constraints on the parameters, upon linearization about an AdS background, yields a unitary and ghost free theory \cite{Bergshoeff:2009aq}. Furthermore, a holographic renormalization study of NMG was conducted in \cite{Alishahiha:2010bw}, where in the context of AdS/CFT correspondence it was suggested that the dual CFT could be a logarithmic conformal field theory (LCFT). Several exact solutions of NMG and their properties can be found in \cite{Arvanitakis:2014yja, Alishahiha:2014dma, Giribet:2014wla, Arvanitakis:2015yya, Deger:2015wpa, Charyyev:2017uuu,Sarioglu:2018rhl,  Sarioglu:2019uxv}.

Recently, a three dimensional stationary black hole solution of NMG, called stationary Lifshitz black hole\footnote{The name ``stationary
	Lifshitz black hole'' was depicted by the author in \cite{Sarioglu:2018rhl}, since it derives from the static Lifshitz black hole, even though the metric in Eq. (\ref{metric}) is neither left invariant under the proper Lifshitz
	scalings, nor asymptotically Lifshitz. }, has been obtained by Sariouglu in \cite{Sarioglu:2018rhl}. It can be derived by performing a specific boost on the static Lifshitz black hole \cite{Cai:2009ac}. In general, only few exact static Lifshitz black hole solutions \cite{AyonBeato:2009nh, Mann:2009yx, Donos:2010tu, Lu:2012xu, Catalan:2014una, Balasubramanian:2009rx, Dehghani:2010kd, AyonBeato:2010tm} in various modified theories of gravity and even less stationary ones \cite{Cai:2009ac} are currently known. As it turns out, such Lifshitz spacetimes play an important role in non-relativistic holography, where studies of critical phenomena in strongly correlated non-relativistic gauge theories at finite temperatures have dual description in terms of Lifshitz gravitational backgrounds \cite{Kachru:2008yh, Balasubramanian:2009rx}. The latter suggests that the thermodynamics of the stationary Lifshitz black hole should be an essential ingredient for understanding the properties of its quantum dual. This motivates us to study the phase structure and the thermodynamics of the new stationary Lifshitz black hole  solution, which is the main goal of this paper. For this purpose, our investigation will take advantage of some known standard and non-standard statistical and purely geometrical techniques.

A special class of non-standard tools for studying the equilibrium thermodynamics of gravitational systems falls within the formalism of the so called Thermodynamic Information Geometry. It utilizes powerful concepts from differential geometry and mathematical statistics, thus making it a very useful framework. This is due to the fact that geometry studies mutual relations between elements, such as distance and curvature, thus one can naturally uncover essential features and gain valuable insights of the system under consideration. Thermodynamic geometry was first introduced by Weinhold \cite{doi:10.1063/1.431689} in 1975 and later by Ruppeiner \cite{RevModPhys.67.605}. Weinhold showed that the  laws of equilibrium thermodynamics can be represented in terms of an abstract metric space. This can be achieved by utilizing the Hessian of the internal energy with respect to the extensive parameters of the system and considering it as a Riemannian metric on the space of macro states. On the other hand, Ruppeiner developed his geometric approach within fluctuation theory, where one implements the entropy as a thermodynamic potential. Here, one can use the Hessian of the entropy to find the probability for fluctuation between different macro states. Later it was discovered that both metric approaches are conformally related via the temperature being the conformal factor.

However, Hessian thermodynamics is not the only way to define a Riemannian metric on the equilibrium manifold. More general approach was proposed by Quevedo in \cite{Quevedo:2017tgz}, who considers Legendre invariant metrics. The latter preserve the physical properties of the system under different choices of thermodynamic potential, but there are infinitely many Legendre invariant metrics to choose from. For this reason, two additional approaches were proposed. The first one is given by Hendi, Panahiyan, Panah and Momennia (HPEM) in \cite{Hendi:2015rja}, where the authors consider thermodynamic metric with specific conformal function, which seems to resolve the problem of redundant singularities in Quevedo's approach. The second one is considered by Mirza and Mansoori (MM) in \cite{Mansoori:2013pna, HosseiniMansoori:2019jcs,Mansoori:2014oia, Mansoori:2016jer}, which is based on conjugate thermodynamic potentials, specifically chosen to reflect the relevant thermodynamic properties of system under consideration. Some applications of these approaches to different gravitational systems can be found for example in \cite{Hendi:2015fya, Hendi:2015cqz, Hendi:2016usw, EslamPanah:2018ums, Aman:2003ug, Shen:2005nu, Cai:1998ep, Aman:2007pp, Sarkar:2006tg, Mansoori:2016jer, Mansoori:2014oia, Ruppeiner:2018pgn, Vetsov:2018dte, Mansoori:2013pna, HosseiniMansoori:2019jcs}. In order to identify the admissible thermodynamic metrics for a given black hole solution, a case by case study is required.

This paper is organized as follows. In Section \ref{sec 2} we present the stationary Lifshitz black hole solution of New Massive Gravity obtained in \cite{Sarioglu:2018rhl}. Here, we calculate some of the curvature invariants of the solution and identify the relevant physical singularities. We also find the location of the event horizon and the Hawking temperature by investigating the Killing symmetries of the solution.Additionally, the Smarr relation between the relevant thermodynamic parameters has also been obtained. In Section \ref{sec 3} we identify the parameter regions of local and global thermodynamic stability of the black hole solution in different ensembles. In Section \ref{sec 4} we study the problem of thermodynamic stability within the framework of Thermodynamic Information Geometry. We identify the admissible thermodynamic metrics and study their properties. In Section \ref{sec 5} we investigate geodesics on the space of equilibrium states, which correspond to the available thermodynamic metrics. This allows us to study the  thermodynamic length (the shortest distance) between two macro states, which can be used to optimize the implementation of quasi-static protocols in a given ensemble. Finally, in Section \ref{sec 6} we give our concluding remarks.

\section{Stationary Lifshitz black hole solution of NMG}\label{sec 2}

The three-dimensional new massive gravity (NMG) was originally proposed in \cite{PhysRevLett.102.201301, Bergshoeff:2009aq} as a parity-preserving and unitary solution to the problem of consistently extending the Fierz-Pauli field theory for a massive spin-2 particle to include interactions. Its action is given by
\begin{equation}
S =  \int d^3 x \sqrt{-g} \Big( R - 2\Lambda_0 + \frac{1}{m^2} \big( S^{\mu \nu}S_{\mu \nu} - S^2 \big) \Big),
\end{equation}
where $\Lambda_0 $ is the cosmological constant, $m$ is a mass parameter, and $S_{\mu \nu}$ is the Shouten tensor,
\begin{equation}
S_{\mu \nu} \equiv R_{\mu \nu} - \frac{1}{4} R g_{\mu \nu},\quad S=g^{\mu \nu} S_{\mu \nu}= \frac{R}{4}.
\end{equation}
The field equations for the metric can be derived by varying the action with respect to the metric tensor, thus
\begin{equation}
\label{eqField}
R_{\mu \nu} - \frac{1}{2} R g_{\mu \nu} + \Lambda_0 g_{\mu \nu} + \frac{1}{m ^2} K_{\mu \nu} = 0,
\end{equation}
where we have defined the following tensor quantity
\begin{equation}
K_{\mu \nu} \equiv \Box S_{\mu \nu} - \nabla_{\mu} \nabla _{\nu} S + S S_{\mu \nu} - 4 S_{\mu \rho} S_{\nu}^{\rho} + \frac{1}{2} g_{\mu \nu} \big( 3 S^{\rho \sigma} S_{\rho \sigma} - S^2 \big).
\end{equation}
Imposing specific choice of the parameters, namely
\begin{equation}
\Lambda_0 = - \frac{13}{2\ell^2},\qquad m^2 = \frac{1}{2\ell^2},
\end{equation}
one can obtain as a solution the static Lifshitz black hole \cite{AyonBeato:2009nh}
\begin{equation}\label{eqSLBH}
d{s^2} =  - {x^3}\left( {1 - \frac{M}{x}} \right)d{t^2} + \frac{{{\ell ^2}d{x^2}}}{{4x(x - M)}} + {\ell ^2} x^2 d{\theta ^2},
\end{equation}
where $x=\rho^2/\ell^2$. Now, boosting the metric (\ref{eqSLBH}) via
\begin{equation}
\left( \begin{array}{l}
dt\\
d\theta 
\end{array} \right) \to \frac{1}{{\sqrt {1 - {\omega ^2}} }}\left( {\begin{array}{*{20}{c}}
1&{ - \omega \ell }\\
{ - \omega/\ell}&1
\end{array}} \right)\left( \begin{array}{l}
dt\\
d\theta 
\end{array} \right),
\end{equation}
where $\omega$ is a real constant with $|\omega|<1$, one arrives at the stationary Lifshitz metric \cite{Sarioglu:2018rhl}:
\begin{align}
\label{metric}
\nonumber 
d{s^2} = g_{\mu\nu}dx^\mu dx^\nu&= - \frac{{x({x^2} - Mx - {\omega ^2})}}{{1 - {\omega ^2}}}d{t^2} + \frac{{2\omega \ell x({x^2} - Mx - 1)}}{{1 - {\omega ^2}}}dtd\theta 
\\
 &+ \frac{{{\ell ^2}}}{{4x(x - M)}}d{x^2} 
 + \frac{{{\ell ^2}x(1 - {\omega ^2}{x^2} + M{\omega ^2}x)}}{{1 - {\omega ^2}}}d{\theta ^2}.
\end{align}
Without loss of generality, we can assume $ 0\leq \omega<1$. Solving the equation $g^{xx}=0$ one finds two positive roots, namely the event horizon at $x=M$, and a central singularity at $x=0$. The root $x=0$ is a true singularity, due to the divergence of the Ricci curvature,
\begin{equation}\label{eqRicciPhysx}
R = \frac{2({4M - 13x})}{{x{\ell ^2}}}.
\end{equation}
Some higher-order invariants also confirm this statement, namely
\begin{equation}\label{eqKrechInv}
{R^{\alpha \beta }}_{\delta \gamma }{R^{\delta \gamma }}_{\alpha \beta } = \frac{4({8{M^2} - 48Mx + 91{x^2}})}{{{x^2}{\ell ^4}}},
\end{equation}
\begin{equation}
{R^{\alpha \beta }}_{\delta \gamma }{R^{\sigma \tau }}_{\alpha \beta }{R^{\delta \gamma }}_{\sigma \tau } =  \frac{8({ 16{M^3} - 144{M^2}x +540M{x^2} - 757{x^3}})}{{{x^3}{\ell ^6}}}.
\end{equation}
\begin{equation}\label{eqKarhInv}
{\nabla_\sigma R_{\alpha \beta \gamma \delta  }}{\nabla^\sigma R^{\alpha \beta \gamma \delta  }} =  \frac{{64(x-M)(4{M^2} - 6Mx + 9{x^2})}}{{{x^3}{\ell ^6}}}.
\end{equation}

Furthermore, considering the Killing symmetries of the solution we can  identify the surface $x=M$ with a Killing horizon $H$. In this case, the standard rotating Killing vector takes the form
\begin{equation}
K^\mu=K_{(t)}^\mu+\Omega K_{(\theta)}^\mu,
\end{equation}
where $K_{(t)}^\mu=(1,0,0)^T$, $K_{(\theta)}^\mu=(0,0,1)^T$, and $\Omega$ is the angular velocity on the horizon. If the surface $x=M$ is a Killing horizon, then the Killing vector should become null on $H$, i.e.
\begin{equation}\label{eqKillingNull}
g_{\mu\nu} K^\mu K^\nu=K_\mu K^\mu=0.
\end{equation}
There are two solutions to this equation, namely
\begin{equation}
{\Omega _ \pm } = \frac{{\omega (Mx - {x^2} + 1) \pm ({\omega ^2} - 1)\sqrt {x(x - M)} }}{{x{\omega ^2}\ell (M - x) + \ell }},
\end{equation}
which, on the surface $x=M$, yield a constant angular velocity
\begin{equation}
\Omega=\frac{\omega}{\ell}.
\end{equation}
Therefore, the rigidity theorem is valid and $x=M$ satisfies all requirements for an event horizon. One can now calculate the Hawking temperature $T$ corresponding to the event horizon $x=M$. In this case, it is proportional to the surface gravity $\kappa$ defined by
\begin{equation}\label{key}
\kappa^2=-\frac{1}{2} (\nabla_\mu K_\nu)(\nabla^\mu K^\nu)|_H
\end{equation}
on the horizon $H$, thus one finds
\begin{eqnarray}\label{eqTandOmega}
T = \frac{\kappa}{2 \pi}=\frac{{{M^{3/2}}\sqrt {1 - {\Omega ^2}{\ell ^2}} }}{{2 \pi \ell }}.
\end{eqnarray}
The thermodynamics of the stationary Lifshitz black hole (\ref{metric}) is further described by the energy $E$, the entropy $S$, and the angular momentum $L$, as found in \cite{Sarioglu:2018rhl}
\begin{equation}
S = \frac{{2\pi \ell \sqrt M }}{{\sqrt {1 - {\Omega ^2\ell^2}} }},\qquad E = \frac{{{M^2}\left( {1+3{\Omega ^2\ell^2} } \right)}}{{4\left( {1 - {\Omega ^2\ell^2}} \right)}},\qquad L = \frac{{{M^2}\Omega \ell^2 }}{{ {1 - {\Omega ^2\ell^2}}}}.
\end{equation}
One can check that the first law of thermodynamics is satisfied
\begin{equation}
\label{firstlaw}
dE=TdS+\Omega dL,
\end{equation}
together with the Smarr relation
\begin{equation}
E=\frac{1}{4}TS+\Omega L.
\end{equation}
The latter is a direct consequence of Euler’s
theorem for quasi-homogeneous function, where we can simply write down $E = E(S,L)$ as
\begin{equation}
E(S,L) = \frac{1}{4}\frac{{\partial E}}{{\partial S}}S + \frac{{\partial E}}{{\partial L}}L.
\end{equation}
Hence, under re-scaling with a parameter $\alpha$, one has $E(\alpha S,\alpha^4 L)=\alpha^4 E(S,L)$, which shows that the energy $E$ is a quasi-homogeneous function of degree 4. This can be directly confirmed by analyzing the roots of the following cubic equation
\begin{equation}
{E^3} - \frac{1}{4}{\left( {\frac{S}{{2\pi \ell }}} \right)^4}{E^2} - \frac{{9{L^2}}}{{8{\ell ^2}}}E + {L^2}\left( {\frac{{27{\pi ^4}{L^2}}}{{4{S^4}}} + \frac{{{S^4}}}{{64{\pi ^4}{\ell ^6}}}} \right) = 0.
\end{equation}
We present this simple, but lengthy calculation in Appendix \ref{appA}.

\section{Local and global thermodynamic stability}
\label{sec 3}
\subsection{Specific heats and local thermodynamic stability}

Throughout the paper the local coordinates on the equilibrium space of macro-states for the stationary Lifshitz black hole are going to be the intensive parameters $T$ and $\Omega$.
Considering $0\leq \omega<1$ and $\ell> 0$, one finds
\begin{equation}\label{eqMetricExistenceCond}
0\leq \Omega \ell<1.
\end{equation}
Solving Eq. (\ref{eqTandOmega}) for $M$ in terms of ($T,\Omega$),
\begin{equation}
M = {\left( {\frac{{2\pi T\ell }}{{\sqrt {1 - {\Omega ^2}{\ell ^2}} }}} \right)^{2/3}},
\end{equation}
one can immediately express the relevant extensive thermodynamic quantities in ($T,\Omega$) space
\begin{equation}\label{eqS}
S =  \frac{T^{1/3} (2\pi\ell) ^{4/3}}{\big( 1 - \ell ^2 \Omega^2 \big)^{2/3}},\quad
E=\frac{ (\pi\ell T )^{4/3} \left(1+3 \Omega ^2 \ell ^2\right)}{2^{2/3} \left(1-\Omega ^2 \ell ^2\right)^{5/3}},\quad
L = \frac{\Omega  \ell ^{2} (2 \pi \ell T)^{4/3}}{\left(1-\Omega ^2 \ell ^2\right)^{5/3}}.
\end{equation}
The specific heats of the black hole in ($T,\Omega$) space are given by \cite{Mansoori:2014oia}
\begin{equation}\label{eqC}
C_\Omega (T,\Omega) = T \left(\frac{\partial S}{\partial T}\right)_\Omega = \frac{T^{1/3} (2\pi\ell) ^{4/3}}{3\big( 1 - \ell ^2 \Omega^2 \big)^{2/3}},
\end{equation}
and 
\begin{equation}\label{eqCL}
{C_L}(T,\Omega ) = T{\left( {\frac{{\partial S}}{{\partial T}}} \right)_L} = T\frac{{\left| {\begin{array}{*{20}{c}}
			{{{({\partial _T}S)}_\Omega }}&{{{({\partial _\Omega }S)}_T}}\\
			{{{({\partial _T}L)}_\Omega }}&{{{({\partial _\Omega }L)}_T}}
			\end{array}} \right|}}{{\left| {\begin{array}{*{20}{c}}
			{{{({\partial _T}T)}_\Omega }}&{{{({\partial _\Omega }T)}_T}}\\
			{{{({\partial _T}L)}_\Omega }}&{{{({\partial _\Omega}L)}_T}}
			\end{array}} \right|}} = \frac{{{{(2\pi \ell )}^{4/3}}{T^{1/3}}(1 - 3{\Omega ^2}{\ell ^2})}}{{{{(1 - {\Omega ^2}{\ell ^2})}^{2/3}}(3 + 7{\Omega ^2}{\ell ^2})}}.
\end{equation}
The Davies critical points are the set of divergences for $C_{\Omega,L}$, namely the spinodal
\begin{equation}\label{eqSpinodal}
\Omega \ell =1,
\end{equation}
where the metric (\ref{metric}) is also singular, thus one has to consider only the case $\Omega\neq1/\ell$. The latter is already assured by Eq. (\ref{eqMetricExistenceCond}). The second spinodal occurs when the specific heat changes its sign ($C_\Omega=C_L=0$). For  $C_\Omega$ this is possible only in the extremal case $T=0$, which is not allowed by the principles of thermodynamics. For $C_L=0$, besides the case $T=0$, the change of sign occurs also on the curve
\begin{equation}
\sqrt 3 \ell \Omega  = 1.
\end{equation}
On the other hand, local thermodynamic stability requires only positive specific heats $C_{\Omega,L}>0$, which leads to ($T>0$, $0\leq \Omega \ell<1$) for $C_\Omega$, and ($T>0$, $0\leq \sqrt 3 \ell \Omega  <1$) for $C_L$. Therefore, if we want for both specific heats to be positive, one has to impose
\begin{equation}\label{eqLTDS}
0\leq \Omega<\frac{1}{\sqrt 3 \ell},\quad \ell>0,\quad T>0.
\end{equation}
This is the condition for local thermodynamic stability of the stationary Lifshitz black hole solution of NMG. However, one can also consider a weaker condition for values of the angular velocity in the range $1/(\sqrt 3 \ell)\leq \Omega<\ell$. In this case, the black hole is locally stable from thermodynamic standpoint only with respect to $C_\Omega$, but not with respect to $C_L$.

\subsection{Ensembles and global thermodynamic stability}
While local thermodynamic stability identifies whether a certain phase in equilibrium is a local maximum of the total entropy, global thermodynamic stability is concerned with phases of the system corresponding to the global maximum. In canonical and grand-canonical ensembles local thermodynamic stability translates to positive specific heats, while global stability under thermal fluctuations translates in the concavity of Helmholtz and Gibbs free energies respectively.

In canonical ensemble the preferred phase of the system is the one that minimizes the Helmholtz free energy \cite{Monteiro:2010cq},
\begin{equation}
F(T,\Omega ) = E - TS =- \frac{{{}{{(\pi T\ell )}^{4/3}}\left( {3-7{\Omega ^2}{\ell ^2}} \right)}}{{{2^{2/3}}{{\left( {1 - {\Omega ^2}{\ell ^2}} \right)}^{5/3}}}}.
\end{equation}
It has local extrema at $\Omega=\sqrt{21}/(7\ell)$ and $\Omega=0$, which are saddle curves for arbitrary values of the temperature $T>0$. The free energy $F$ and its derivatives are discontinuous on $\Omega\ell=1$, thus indicating a phase transition.
In order for the black hole to be in a global thermodynamic equilibrium, the following concavity condition must be satisfied
\begin{equation}
\frac{{{\partial ^2}F}}{{\partial {T^2}}} =  - \frac{{(2 \pi \ell) ^{4/3}\left( {3 - 7{\ell ^2}{\Omega ^2}} \right)}}{{9{T^{2/3}}{{\left( {1 - {\ell ^2}{\Omega ^2}} \right)}^{5/3}}}} < 0.
\end{equation}
This leads to the constraint
\begin{equation}\label{eqFcond}
\Omega  < \frac{{\sqrt {21} }}{{7\ell }},
\end{equation}
which is less restrictive than the condition (\ref{eqLTDS}) for local thermodynamic stability.

In the grand-canonical ensemble the preferred phase of the system is the one that minimizes the Gibbs free energy,
\begin{equation}
G(T,\Omega ) = E - TS - \Omega L =  - \frac{{3{}}}{{{2^{2/3}}}}{\left( {\frac{{\pi T\ell }}{{\sqrt {1 - {\ell ^2}{\Omega ^2}} }}} \right)^{4/3}}.
\end{equation}
Considering $T>0$, the local extremum is located at $\Omega=0$, which is a local maximum. The Gibbs free energy and its derivatives are also discontinuous at $\Omega\ell=1$. The concavity condition in this case,

\begin{equation}
\frac{{{\partial ^2}G}}{{\partial {T^2}}} = -\frac{(2 \pi \ell) ^{4/3}}{3 T^{2/3}\left(1-\Omega ^2 \ell ^2\right)^{2/3}}<0,
\end{equation}
assures global thermodynamic stability. This is always true within the range given in Eq. (\ref{eqMetricExistenceCond}).

One notes that the conditions for global thermodynamic stability in both ensembles restricts distinctively the angular velocity $\Omega$ of the black hole. However, the condition for local thermodynamic stability (\ref{eqLTDS}) always falls within them. 

The thermodynamic stability of the stationary Lifshitz black hole solution can be further analyzed by identifying the proper Riemannian metrics on the space of equilibrium states, which we show in the next section.

\section{Thermodynamic geometry on the equilibrium manifold}\label{sec 4}

In this section we investigate the thermodynamic stability of the system by several Riemannian metrics defined on ($T,\Omega$) equilibrium space of the stationary Lifshitz black hole. 

\subsection{Hessian thermodynamic metrics}

The simplest choice one can consider is the Ruppeiner metric defined by the Hessian of the entropy,
  \begin{equation}\label{eqRuppeinerMetric}
   g^{(R)}_{ab}=-\partial_a\partial_b S(T,\Omega)=\left( {\begin{array}{*{20}{c}}
{\frac{{2{{(2\pi \ell )}^{4/3}}}}{{9{T^{5/3}}{{(1 - {\ell ^2}{\Omega ^2})}^{2/3}}}}}&{ - \frac{{4\Omega {\ell ^2}{{(2\pi \ell )}^{4/3}}}}{{9{T^{2/3}}{{(1 - {\ell ^2}{\Omega ^2})}^{5/3}}}}}\\
\vspace{-0.2cm}&\vspace{-0.2cm}\\
{ - \frac{{4\Omega {\ell ^2}{{(2\pi \ell )}^{4/3}}}}{{9{T^{2/3}}{{(1 - {\ell ^2}{\Omega ^2})}^{5/3}}}}}&{ - \frac{{4{\ell ^2}{{(2\pi \ell )}^{4/3}}{T^{1/3}}\left( {3 + 7{\ell ^2}{\Omega ^2}} \right)}}{{9{{\left( {1 - {\ell ^2}{\Omega ^2}} \right)}^{8/3}}}}}
\end{array}} \right),
  \end{equation}
where $\partial _a$ denotes derivatives with respect to ($T, \Omega$). Due to the probabilistic interpretation of Hessian metrics \cite{PhysRevA.20.1608}, one requires their positive definiteness. This can be assured by imposing Sylvester's criterion, which states that all the principal minors of the metric tensor be strictly positive definite, i.e.
\begin{equation}\label{eqSylvCritGeneral2d}
g_{TT} > 0,\quad g_{\Omega \Omega } > 0,\quad \det(g_{ab}) > 0.
\end{equation}
Unfortunately, these conditions cannot be  simultaneously satisfied for $g^{(R)}_{ab}$, because $g^{(R)}_{\Omega \Omega }<0$ is always negative, thus there are no sub-regions in ($T,\Omega$) space, where the metric tensor (\ref{eqRuppeinerMetric}) is positive definite. The same is true if one considers the angular momentum $L$  or the Helmholtz free energy $F$ as thermodynamic potentials in ($T,\Omega$) space. Therefore, we will not study these metrics here.

On the other hand, one can take advantage of Weinhold's approach utilizing the Hessian of the internal energy of the system instead of the entropy \cite{Weinhold:1975}. In this case, one finds
 \begin{equation}\label{eqWeinholdMetric}
  g^{(W)}_{ab}=\partial _a \partial _ b E(T,\Omega)=\left(
  \begin{array}{cc}
  \frac{(2\pi \ell) ^{4/3} (3 \Omega ^2 \ell ^2+1)}{9 T^{2/3} \left(1-\ell ^2 \Omega ^2\right)^{5/3}} & \frac{4\Omega (2 \pi\ell) ^{4/3} T^{1/3}   (3 \ell ^2 \Omega ^2+7)}{9 \left(1-\ell ^2 \Omega ^2\right)^{8/3}} \\
  \vspace{-0.2cm}&\vspace{-0.2cm}\\
  \frac{4 \Omega(2 \pi\ell) ^{4/3} T^{1/3} (3 \ell ^2 \Omega ^2+7)}{9 \left(1-\ell ^2 \Omega ^2\right)^{8/3}} & \frac{\ell ^2 (2\pi T \ell )^{4/3} \left(21 \ell ^4 \Omega ^4+118 \ell ^2 \Omega ^2+21\right)}{9 \left(1-\ell ^2 \Omega ^2\right)^{11/3}} \\
  \end{array}
  \right) .
  \end{equation}
  \normalsize
Here, the Sylvester criterion (\ref{eqSylvCritGeneral2d}) is applicable and further restricts the possible values of the angular velocity. To show this, one notes that the first two principal minors $g^{(W)}_{TT}$ and  $g^{(W)}_{\Omega\Omega}$ of the metric are strictly positive only for $\ell \Omega  < 1$. On the other hand, the third principal minor, i.e. the determinant of the metric, is given by
\begin{equation}\label{eqWdet}
\det {{\hat g}^{(W)}} =  - \frac{{{\ell ^2}{T^{2/3}}{{(2\pi \ell )}^{8/3}}(27{\ell ^6}{r^3} + 99{\ell ^4}{r^2} + 201{\ell ^2}r - 7)}}{{27{{(1 - {\ell ^2}r)}^{16/3}}}},
\end{equation}
where we have substituted $\Omega^2=r$. The expression in Eq. (\ref{eqWdet}) is strictly positive for $r<r_{+}$, where $r_{+}\approx 0.0342/\ell^2$ is the only positive real root of the cubic expression   
\begin{equation}\label{eqCubicOmega}
27{\ell ^6}{r^3} + 99{\ell ^4}{r^2} + 201{\ell ^2}r - 7=0.
\end{equation}
In terms of the angular velocity $\Omega$ one finds
\begin{equation}\label{eqAngVelWeinhold}
0 \leq \Omega  < \sqrt r_{+} ,
\end{equation}
which is more restrictive than $\ell \Omega <1$. The Weinhold scalar curvature yields
  \begin{equation}
  R^{(W)}(T,\Omega)= \frac{3\times 2^{2/3} \left(1-\Omega ^2 \ell ^2\right)^{8/3} \left(9 \Omega ^2 \ell ^2 \left(9 \Omega ^4 \ell ^4-57 \Omega ^2 \ell ^2+7\right)+49\right)}{ (\pi T \ell )^{4/3} \left(27 \Omega ^6 \ell ^6+99 \Omega ^4 \ell ^4+201 \Omega ^2 \ell ^2-7\right)^2}.
  \end{equation}
It is singular at the root $\Omega=\sqrt{r_{+}}$, but this point is safely excluded by Sylvester's criterion. In other words, no geodesics (quasi-static processes) can pass through this spinodal within this approach. Moreover, the root $\sqrt r_{+}\approx 0.185/\ell$ lies within the region of local thermodynamic stability (\ref{eqLTDS}), suggesting  Weinhold's metric as a viable metric in the ($T,\Omega$) space.

We can now follow the standard interpretation \cite{2010AmJPh..78.1170R}, where the sign of the thermodynamic scalar curvature can be linked to the
nature of the inter-particle interactions in composite thermodynamic systems. In this case, we find that $R^{(W)}>0$ within the region given in Eq. (\ref{eqAngVelWeinhold}), which suggest repulsive interactions in the gravitational theory and hence in the dual gauge theory. Moreover, looking at the following limits
\begin{align}
&\mathop {\lim }\limits_{\ell  \to \infty } {R^{(W)}}\big |_{T=const} = 0,\quad \mathop {\lim }\limits_{T \to \infty } {R^{(W)}}\big |_{\ell=const} = 0,
\\
&\mathop {\lim }\limits_{\Omega  \to 1/\ell } {R^{(W)}}\big |_{T,\ell=const} = 0,\quad \mathop {\lim }\limits_{\Omega  \to 0} {R^{(W)}}\big |_{T,\ell=const} = \frac{{3 \times {2^{2/3}}}}{{{{(\pi T\ell )}^{4/3}}}},
\end{align}
one finds that for large $\ell$ at a fixed temperature the Weinhold thermodynamic curvature vanishes, thus the correlations between the particles become weak and we approach free non-interacting system. The same is true for large temperatures at fixed $\ell$. For states near the curve $\Omega\ell=1$, the curvature $R^{(W)}$ is also vanishing, thus we have a weakly coupled system near $\Omega\ell=1$. In the static case, $\Omega=0$, the strength of the interactions saturates at a value inversely proportional to the temperature $T$ of the black hole. Therefore, in this case, the interactions weaken for larger temperatures and strengthened for smaller temperatures.

One can consider also the Hessian of the Gibbs free energy
\begin{equation}\label{eqHessGibbsMetric}
g_{ab}^{(G)} =  - {\partial _a}{\partial _b}G(T,\Omega ) = \left( {\begin{array}{*{20}{c}}
{\frac{{{{(2\pi \ell )}^{4/3}}}}{{3{T^{2/3}}{{\left( {1 - {\ell ^2}{\Omega ^2}} \right)}^{2/3}}}}}&{\frac{{4\Omega {\ell ^2}{{(2\pi \ell )}^{4/3}}{T^{1/3}}}}{{3{{\left( {1 - {\ell ^2}{\Omega ^2}} \right)}^{5/3}}}}}\\
  \vspace{-0.2cm}&\vspace{-0.2cm}\\
{\frac{{4\Omega {\ell ^2}{{(2\pi \ell )}^{4/3}}{T^{1/3}}}}{{3{{\left( {1 - {\ell ^2}{\Omega ^2}} \right)}^{5/3}}}}}&{\frac{{{\ell ^2}{{(2\pi \ell T)}^{4/3}}\left( {3 + 7{\ell ^2}{\Omega ^2}} \right)}}{{3{{\left( {1 - {\ell ^2}{\Omega ^2}} \right)}^{8/3}}}}}
\end{array}} \right).
\end{equation}
Imposing Sylvester's criterion one finds
\begin{equation}\label{eqSylvCritGibbs}
0 \leq \Omega  < \frac{{1 }}{{\sqrt 3\ell }},
\end{equation}
which precisely coincides with the range for local thermodynamic stability (\ref{eqLTDS}), thus one can take the Gibbs metric as a viable thermodynamic metric. The upper limit $\Omega=\sqrt{3}/(3\ell)$ is where the inverse of the Gibbs metric becomes singular. The scalar curvature is given by
\begin{equation}
{R^{(G)}}(T,\Omega ) = \frac{{{2^{2/3}}{{(1 - {\ell ^2}{\Omega ^2})}^{5/3}}}}{{{\pi ^{4/3}}{T^{4/3}}{\ell ^{4/3}}{{(1 - 3{\ell ^2}{\Omega ^2})}^2}}},
\end{equation}
which is positive in the range (\ref{eqSylvCritGibbs}), specified by Sylvester's criterion. This suggests elliptic information geometry, which corresponds to repulsive inter-particle interactions. As in the Weinhold's case, the Gibbs curvature vanishes for high temperatures or large values of the gravitational parameter $\ell$. In the static case $R^{(G)}$ is again inversely proportional to the temperature
\begin{equation}
\mathop {\lim }\limits_{\Omega  \to 0} {R^{(G)}} = \frac{{{2^{2/3}}}}{{{{(\pi T\ell )}^{4/3}}}}.
\end{equation}

In this subsection we have considered the most commonly used  Hessian thermodynamic metrics. In what follows, we are going to consider the New Thermodynamic Geometry approach to the equilibrium manifold of the Lifshitz black hole.

\subsection{New thermodynamic geometry}

Although their convenient probabilistic interpretation Hessian thermodynamic metrics often fails to reproduce all relevant critical points. One way to avoid such problem is proposed in \cite{Mansoori:2013pna, HosseiniMansoori:2019jcs,Mansoori:2014oia, Mansoori:2016jer}, where the authors take advantage of conjugate thermodynamic potentials to construct the proper Riemannian metrics on the space of equilibrium states of a given black hole system. 
Within the formalism of the New Thermodynamic Geometry (NTG) \cite{HosseiniMansoori:2019jcs} one can find a positive definite metric on ($T,\Omega$) space by utilizing the Gibbs free energy as conjugate potential. For this purpose, let us take the differential from both sides of $G=E-TS-\Omega L$,
\begin{equation}
dG = dE - TdS - SdT - \Omega dL - Ld\Omega .
\end{equation}
Now, we can express $dE$ and use the first law (\ref{firstlaw}) to find
\begin{equation}\label{eqGibbsFirstLaw}
dG(T,\Omega) =  - SdT - Ld\Omega .
\end{equation}
The metric with respect to $G(T,\Omega)$ is defined by
\begin{equation}\label{eqGibbsTDmetricMM}
 \tilde g_{ab}^{(G)} =  - \frac{1}{T}{\partial _a}{\partial _b}G(T,\Omega ) = \left( {\begin{array}{*{20}{c}}
{\frac{{2\sqrt[3]{2}{\pi ^{4/3}}{\ell ^2}}}{{3T{{\left( {T\left( {\ell  - {\ell ^3}{\Omega ^2}} \right)} \right)}^{2/3}}}}}&{\frac{{8\sqrt[3]{2}{\pi ^{4/3}}{\ell ^4}\Omega }}{{3{{(T\ell )}^{2/3}}{{\left( {1 - {\ell ^2}{\Omega ^2}} \right)}^{5/3}}}}}\\
\vspace{-0.2cm}&\vspace{-0.2cm}\\
{\frac{{8\sqrt[3]{2}{\pi ^{4/3}}{\ell ^4}\Omega }}{{3{{(T\ell )}^{2/3}}{{\left( {1 - {\ell ^2}{\Omega ^2}} \right)}^{5/3}}}}}&{\frac{{2\sqrt[3]{2}{\pi ^{4/3}}{\ell ^3}\sqrt[3]{{T\ell }}\left( {3 + 7{\ell ^2}{\Omega ^2}} \right)}}{{3{{\left( {1 - {\ell ^2}{\Omega ^2}} \right)}^{8/3}}}}}
\end{array}} \right).
\end{equation}
Sylvester's criterion leads to the same restriction as in Eq. (\ref{eqSylvCritGibbs}). However, most unexpectedly, this geometry is Ricci flat everywhere, 
\begin{equation}\label{eqMMGibsR}
 \tilde R^{(G)}(T,\Omega)=\frac{{{2^{2/3}}\left( {4(1 - 4Y){\Omega ^2}{\ell ^2} - 7{\Omega ^4}{\ell ^4} + 3} \right)}}{{9{\pi ^{4/3}}\sqrt[3]{T}{\ell ^{4/3}}{{\left( {3{\Omega ^2}{\ell ^2} - 1} \right)}^3}}}(Y - Y) = 0,
\end{equation}
where $Y = 1 - {\ell ^2}{\Omega ^2}$, which suggests free dual gauge theory. In order to make this more explicit we change coordinates to ($S,\Omega$) space. Taking into account the Jacobian of the transformation,
\begin{equation}
J = \frac{{\partial (T,\Omega )}}{{\partial (S,\Omega )}} = \left( {\begin{array}{*{20}{c}}
	{{\partial _S}T}&{{\partial _\Omega }T}\\
	{{\partial _S}\Omega }&{{\partial _\Omega }\Omega }
	\end{array}} \right) = \left( {\begin{array}{*{20}{c}}
	{\frac{{3{S^2}{{({\ell ^2}{\Omega ^2} - 1)}^2}}}{{16{\pi ^4}{\ell ^4}}}}&{\frac{{{S^3}\Omega ({\ell ^2}{\Omega ^2} - 1)}}{{4{\pi ^4}{\ell ^2}}}}\\
	0&1
	\end{array}} \right),
\end{equation}
where
\begin{equation}
T = \frac{{{S^3}{{\left( {{\Omega ^2}{\ell ^2} - 1} \right)}^2}}}{{16{\pi ^4}{\ell ^4}}},
\end{equation}
 and $\hat g=J^t.\tilde g^{(G)}.J$, we end up with the metric on the cylinder
\begin{equation}\label{eqIsomEquilMetricFlat}
d\hat s^2 = \frac{3}{S}d{S^2} + \frac{{S{\ell ^2}(1 - 3{\Omega ^2}{\ell ^2})}}{{{{(1-{\Omega ^2}{\ell ^2} )}^2}}}d{\Omega ^2} = d{\sigma ^2} + {\sigma ^2}d{\chi ^2}.
\end{equation}
In the last step we traded our coordinates for
\begin{equation}\label{eqJustBelow419}
\sigma  = 2\sqrt {3S} ,\quad \chi  = \frac{1}{{\sqrt 6 }}\arctan \left( {\frac{{\sqrt 2 \Omega \ell }}{{\sqrt {1 - 3{\Omega ^2}{\ell ^2}} }}} \right) - \frac{1}{2}\arcsin (\sqrt 3 \Omega \ell ).
\end{equation}
Further substitution by $X = \sigma \cos \chi$ and $Y = \sigma \sin \chi $ yields the desired flat two-dimensional Euclidean space. The latter means that the curve $\Omega\ell=1$ is only a coordinate singularity in ($T,\Omega$) space. It is most obvious, if one calculates the relevant thermodynamic quantities in ($S,\Omega$) space
\begin{equation}
E = \frac{{{S^4}(1 - {\ell ^2}{\Omega ^2})(1 + 3{\ell ^2}{\Omega ^2})}}{{64{\pi ^4}{\ell ^4}}},\quad L = \frac{{{S^4}\Omega (1 - {\ell ^2}{\Omega ^2})}}{{16{\pi ^4}{\ell ^2}}},\quad T = \frac{{{S^3}{{(1 - {\ell ^2}{\Omega ^2})}^2}}}{{16{\pi ^4}{\ell ^4}}},
\end{equation}
\begin{equation}
F = \frac{{{S^4}(1 - {\ell ^2}{\Omega ^2})(7{\ell ^2}{\Omega ^2} - 3)}}{{64{\pi ^4}{\ell ^4}}},\quad G =  - \frac{{3{S^4}{{(1 - {\ell ^2}{\Omega ^2})}^2}}}{{64{\pi ^4}{\ell ^4}}},
\end{equation}
and the specific heats of the black hole
\begin{equation}
{C_\Omega }(S,\Omega ) = T{\left( {\frac{{\partial S}}{{\partial T}}} \right)_\Omega } = T\frac{{{{\{ S,\Omega \} }_{S,\Omega }}}}{{{{\{ T,\Omega \} }_{S,\Omega }}}} = T\frac{1}{{{{\left( {\frac{{\partial T}}{{\partial S}}} \right)}_\Omega }}} = \frac{S}{3},
\end{equation}
\begin{equation}
{C_L}(S,\Omega ) = T{\left( {\frac{{\partial S}}{{\partial T}}} \right)_L} = T\frac{{{{\{ S,L\} }_{S,\Omega }}}}{{{{\{ T,L\} }_{S,\Omega }}}} = T\frac{{\left| {\begin{array}{*{20}{c}}
			{{{({\partial _S}S)}_\Omega }}&{{{({\partial _\Omega }S)}_S}}\\
			{{{({\partial _S}L)}_\Omega }}&{{{({\partial _\Omega }L)}_S}}
			\end{array}} \right|}}{{\left| {\begin{array}{*{20}{c}}
			{{{({\partial _S}T)}_\Omega }}&{{{({\partial _\Omega }T)}_S}}\\
			{{{({\partial _S}L)}_\Omega }}&{{{({\partial _\Omega }L)}_S}}
			\end{array}} \right|}} = \frac{{S(1 - 3{\Omega ^2}{\ell ^2})}}{{3 + 7{\Omega ^2}{\ell ^2}}},
\end{equation}
which are regular everywhere.

Nevertheless, the metric $\tilde g_{ab}^{(G)}$ comes from more general microscopic considerations. Consider a physical system in equilibrium with a large thermal reservoir. The configurational probability distribution is given by
\begin{equation}\label{eqProbDist}
p(y|\lambda ) = \frac{1}{Z}{e^{ - \beta H(y,\lambda )}} = \frac{1}{Z}{e^{ - {\lambda ^i}(t){X_i}(y)}},
\end{equation}
where $y$ is the configuration (a set of random variables or a sample space), $t$ is a time variable, $\beta = 1/T$ is the inverse temperature of the environment ($k_B=1$), $Z$ is the partition function, and $H$ is the Hamiltonian of the system. The Hamiltonian is split into two parts -- collective variables $X_a$ and their conjugate momenta $\lambda^a$, i.e. $\beta H = \lambda^a(t)X_a (y)$. The $\lambda$’s are the experimentally controllable parameters
of the system and define the accessible thermodynamic
state space. In our case, one has $\lambda^a=(T,\Omega)$ and $X_a=(S,L)$. If the partition function $Z$, which normalizes the probability distribution (\ref{eqProbDist}), is calculated in the fixed-$\Omega$ ensemble, then it can be directly related to the Gibbs potential $G$ via \cite{hill1956statistical}
\begin{equation}
\ln Z =  - \beta G = \psi,
\end{equation}
where $\psi$ is the free entropy. In statistical quantum thermodynamics the first derivatives of the free
entropy give the first moments of the collective variables \cite{PhysRevLett.99.100602}
\begin{equation}
\frac{{\partial \psi }}{{\partial {\lambda ^a}}} =  - \left\langle {{X_a}} \right\rangle ,
\end{equation}
while the second derivative yields the covariance matrix
\begin{equation}\label{eqCovMatrix}
{\mathcal{G}_{ab}} = \frac{{{\partial ^2}\psi }}{{\partial {\lambda ^a}\partial {\lambda ^b}}} = \left\langle {\left( {{X_a} - \left\langle {{X_a}} \right\rangle } \right)\left( {{X_b} - \left\langle {{X_b}} \right\rangle } \right)} \right\rangle .
\end{equation}
Substituting $\lambda^a=(T,\Omega)$, $X_a=(S,L)$ and $\beta=1/T$, we find that the covariance matrix (\ref{eqCovMatrix}) exactly corresponds to the thermodynamic metric from Eq. (\ref{eqGibbsTDmetricMM}). Thus the Mirza-Mansoori (MM) approach has a direct relation to quantum statistics.

Let us briefly comment on the results of this section. We have considered the Hessian metric approach to the space of macrostates of the stationary Lifshitz black hole. Here, a case by case study has revealed that a set of viable thermodynamic metrics can be given by the Hessian of the energy $E$ (Weinhold's approach) of the system and the Gibbs free energy $G$. However, the Hessian of the entropy $S$ did not produced a suitable metric formalism on the ($T,\Omega$) thermodynamic manifold. A further investigation lead us to consider the New Thermodynamic Geometry approach, where we have constructed a proper positive definite thermodynamic metric utilizing the Gibbs free energy. It turned out that the new metric lead to a flat statistical manifold, which corresponds to a free non-interacting underlying theory. We have also showed that the new Gibbs metric coincides with the covariance matrix from quantum thermodynamics in the so called fixed-$\Omega$ ensemble.

Next, we are going to consider the optimal (with minimal energy loss) paths on the equilibrium manifold for implementing quasi-static processes, under which the systems has enough time to equilibrate on every basic step. 

\section{Thermodynamic length and quasi-static processes}\label{sec 5}
In this section we consider geodesics on the equilibrium state space spanned by ($T,\Omega$). 
The action for the thermodynamic geodesics is written by \cite{PhysRevLett.99.100602}
\begin{equation}\label{eqGeodAction}
{\cal L} = \int \nolimits_{t_i}^{t_f}  \sqrt {{g_{ab}}(\vec\lambda)\frac{{d{\lambda ^a}}}{{dt}}\frac{{d{\lambda ^b}}}{{dt}}} dt,
\end{equation}
where $t$ is an affine parameter on the geodesics (not necessarily corresponding to time), $\lambda^a(t)=(T(t),\Omega(t))$ are the set of intensive thermodynamic parameters, and ($t_i$, $t_f$) denote the initial and final states.
We can vary the action to obtain the system of coupled geodesic equations
\begin{equation}\label{eqGeodsSys}
{{\ddot \lambda }^c}(t) + \Gamma _{ab}^c(\hat g){{\dot \lambda }^a(t)}{{\dot \lambda }^b(t)} = 0,
\end{equation}
where the dot is a derivative with respect to $t$. By definition the thermodynamic length $\mathcal{L}$, between two equilibrium states at $t_i$ and $t_f$ respectively, is the on-shell value of the action (\ref{eqGeodAction}) for the geodesic curve connecting those states. We can also define a related quantity, called the thermodynamic divergence of the path,
\begin{equation}
{\mathcal {J}} = \tau \int \nolimits_0^\tau  {g_{ab}}(\vec\lambda)\frac{{d{\lambda ^a}}}{{dt}}\frac{{d{\lambda ^b}}}{{dt}}dt,
\end{equation}
which is a measure of the energy dissipation or entropy production for a transition between two equilibrium points  at particular rates of change. In other words, $\mathcal{J}$ measures the efficiency of the quasi-static protocols and satisfies the following bound
\begin{equation}
{\cal J} \ge {{\cal L}^2}.
\end{equation}
The latter follows from the Cauchy-Schwarz inequality for integrals and provides a formal definition of the degree of irreversibility of the process\footnote{With reversibility only for $\mathcal{J}=0$.} (see \cite{RNBA2008} and references therein). We are now ready to begin our analysis of the thermodynamic geodesics in ($T,\Omega$) parameter space.

For the Gibbs metric (\ref{eqGibbsTDmetricMM}) one finds the following coupled geodesic equations
\begin{equation}\label{eqOmegaGeodesicsGibbs}
\ddot \Omega  + \frac{{\dot \Omega }}{3}\left( {\frac{{\dot T}}{T} + \frac{{{\ell ^2}\Omega \dot \Omega (1 - 21{\ell ^2}{\Omega ^2})}}{{3{\ell ^4}{\Omega ^4} - 4{\ell ^2}{\Omega ^2} + 1}}} \right) = 0,
\end{equation}
\begin{equation}\label{eqTempGeodesicsGibbs}
\ddot T - \frac{{5{{\dot T}^2}}}{{6T}} + \frac{{{\ell ^2}(21{\ell ^4}{\Omega ^4} - 22{\ell ^2}{\Omega ^2} + 21)}}{{6(1 - 3{\ell ^2}{\Omega ^2}){{(1 - {\ell ^2}{\Omega ^2})}^2}}}T{{\dot \Omega }^2} = 0,
\end{equation}
where
\begin{equation}\label{eqWsingoftheMetric}
\Omega  \ne \left\{ {\frac{1}{\ell },\frac{{\sqrt 3 }}{{3\ell }}} \right\},\quad T>0.
\end{equation}
The singular points in Eq.  (\ref{eqWsingoftheMetric}) correspond to the singularities of the metric (\ref{eqGibbsTDmetricMM}) and its inverse.
One way to find analytically a non-trivial solution is to consider a quasi-static process with a constant geodesic profile for the angular velocity, $\Omega(t)=\Omega_0=const$. In this case, Eq. (\ref{eqOmegaGeodesicsGibbs}) is trivially satisfied, while Eq. (\ref{eqTempGeodesicsGibbs}) for the temperature becomes simply
\begin{equation}\label{eqTempProfileGibbsMM}
T\ddot T - \frac{5}{6}{{\dot T}^2}=0.
\end{equation}
For later convenience let us solve the more general equation
\begin{equation}\label{eqGenTempEq}
T \ddot T+\alpha \dot T^2=0,
\end{equation}
where $\alpha$ is a real constant. This equation is equvalent to
\begin{equation}
\frac{d}{dt}(T^{\alpha}\dot T)=0,
\end{equation}
giving the first integral
\begin{equation}
T^{\alpha}\dot T=C_1,
\end{equation}
and consequently the general solution to (\ref{eqGenTempEq}):
\begin{equation}
T(t)=\left((\alpha+1)C_1 t+C_2\right)^{\frac{1}{\alpha+1}}.
\end{equation}
Given some initial conditions $T(0)=T_0$ and $\dot{T}(0)={\cal T}_0$, one finds
\begin{equation}
T(t) = T_0^{}{\left( {1 + t(\alpha  + 1)\frac{{{{\cal T}_0}}}{{T_0^{}}}} \right)^{\frac{1}{{\alpha  + 1}}}}.
\end{equation}
In Eq. (\ref{eqTempProfileGibbsMM}) one has $\alpha=-5/6$, thus the temperature profile along the geodesics is
\begin{equation}
T(t) = {T_0}{\left( {1 + t\frac{{{{\cal T}_0}}}{{6{T_0}}}} \right)^6}.
\end{equation}
Here, we have assumed an initial macro-state $T(0)=T_0$ and an initial rate of temperature change $\dot{T}(0)={\cal T}_0$.
The thermodynamic length $\tilde {\cal L}^{(G)}$, between two macro-states at $t=0$ and at $t=\tau$, yields
\begin{equation}
\tilde {\cal L}^{(G)} = \int \nolimits_0^\tau  \sqrt {\tilde g_{TT}^{(G)}\left( {T(t),\Omega (t)} \right){{\dot T}^2}(t)} dt = \frac{{{{(2\pi \ell )}^{2/3}}{|{\cal T}_0}|\tau}}{{\sqrt 3 {\rm{ }}T_0^{5/6}{{(1 - \Omega _0^2{\ell ^2})}^{1/3}}}} .
\end{equation}
For this specific choice of geodesics, the  thermodynamic divergence ${\cal{J}}=\mathcal{L}^2$ saturates the equality, thus $\mathcal{L}^2$ measures also the energy dissipation along the path. In the limit $\Omega_0\to 1/\ell$ the thermodynamic length $\tilde {\cal L}^{(G)}$ becomes infinite. This means that one cannot use ($T,\Omega$) coordinates for quasi-static evolution near this limit. In this case, a change of experimentally controllable parameters of the system, for example to ($X,Y$) chart given just below Eq. (\ref{eqJustBelow419}), is necessary to define meaningful thermodynamic states near $\Omega \ell=1$. The price to pay is that we loose clarity of the physical process, due to the fact that it is unclear how to treat ($X,Y$) as thermodynamic quantities.

When one considers a constant angular velocity geodesic profile for the Hessian-Gibbs metric (\ref{eqHessGibbsMetric}) the system of geodesic equations (\ref{eqGeodsSys}) reduces to
\begin{equation}
\frac{{{\Omega _0}(1 - \Omega _0^2{\ell ^2})}}{{1 - 3\Omega _0^2{\ell ^2}}}\frac{{\dot T}}{T} = 0,\qquad 3\ddot T - \frac{{1 + 5\Omega _0^2{\ell ^2}}}{{1 - 3\Omega _0^2{\ell ^2}}}\frac{{{{\dot T}^2}}}{T} = 0.
\end{equation}
In this case, one has two options, namely $\Omega_0=0$ or $\Omega_0=1/\ell$. The first option, $\Omega(t)=\Omega_0=0$, leads to
\begin{equation}\label{eqTempGeodsGibbsMM0}
3T\ddot T = {{\dot T}^2},
\end{equation}
with the following solution
\begin{equation}\label{eqTempSolGeodsGibbsMM0}
T(t) = {T_0}{\left( {1 + t\frac{{2{{\cal T}_0}}}{{3{T_0}}}} \right)^{3/2}}.
\end{equation}
Therefore, the corresponding thermodynamic length reduces to
\begin{equation}
{{\cal L}^{(G)}} = \int_0^\tau  {\sqrt {g_{TT}^{(G)}\left( {T(t),\Omega (t)} \right){{\dot T}^2}(t)} } dt = \frac{{{{(2\pi \ell )}^{2/3}}\tau \left| {{{\cal T}_0}} \right|}}{{3T_0^{1/3}}}.
\end{equation}
It is always finite, thus a quasi-static protocol is always possible on this path.
The second choice, $\Omega_0=1/\ell$, leads to non-singular geodesic equation
\begin{equation}
T\ddot T + {{\dot T}^2} = 0
\end{equation}
with solution
\begin{equation}
T(t) = {T_0}{\left( {1 + t\frac{{2{\mathcal{T}_0}}}{{{T_0}}}} \right)^{1/2}}.
\end{equation}
However, in the limit $\Omega\to 1/\ell$, the thermodynamic length diverges
\begin{equation}
{\cal L}_2^{(G)} = \left[ {1 - {{\left( {1 + 2\tau \frac{{{{\cal T}_0}}}{{{T_0}}}} \right)}^{1/3}}} \right]\mathop {\lim }\limits_{\Omega _0^{} \to 1/\ell } \frac{{\sqrt 3 {{(2\pi \ell {T_0})}^{2/3}}}}{{2{{(1 - \Omega _0^2{\ell ^2})}^{1/3}}}} = \infty .
\end{equation}

In Weinhold's case (\ref{eqWeinholdMetric}), if one is moving along a path of constant angular velocity, $\Omega(t)=\Omega_0$, the system of  coupled geodesic equations (\ref{eqGeodsSys}) reduces to
\begin{equation}\label{eqWeinholdOmegaGeodsZero}
\frac{{{\Omega _0}(1 - \Omega _0^2{\ell ^2})(9\Omega _0^4{\ell ^4} + 24\Omega _0^2{\ell ^2} + 7)}}{{27\Omega _0^6{\ell ^6} + 99\Omega _0^4{\ell ^4} + 201\Omega _0^2{\ell ^2} - 7}}\frac{{\dot T}}{T} = 0,
\end{equation}
\begin{equation}\label{eqWeinholdTempGeodsZero}
3\ddot T + \frac{{45\Omega _0^6{\ell ^6} + 237\Omega _0^4{\ell ^4} + 191\Omega _0^2{\ell ^2} + 7}}{{27\Omega _0^6{\ell ^6} + 99\Omega _0^4{\ell ^4} + 201\Omega _0^2{\ell ^2} - 7}}\frac{{{{\dot T}^2}}}{T} = 0.
\end{equation}
A non-trivial profile for the temperature $T(t)$ can be obtained if we set $\Omega_0=0$. In this case, the second equation (\ref{eqWeinholdTempGeodsZero}) turns out to be the same as in the previous case (\ref{eqTempGeodsGibbsMM0}) with the same solution (\ref{eqTempSolGeodsGibbsMM0}) for $T(t)$. However, the thermodynamic length of the path, connecting two macro states, is different
\begin{equation}
{{\cal L}^{(W)}} = \int_0^\tau  {\sqrt {g_{TT}^{(W)}\left( {T(t),\Omega (t)} \right){{\dot T}^2}(t)} } dt = \frac{{{{(2\pi \ell )}^{2/3}}\tau \left| {{{\cal T}_0}} \right|}}{{{{(3{T_0})}^{1/3}}}}.
\end{equation}
The other possibility in Eqs. (\ref{eqWeinholdOmegaGeodsZero}) and (\ref{eqWeinholdTempGeodsZero}) is $\Omega=1/\ell$, where the thermodynamic length is again singular
\begin{equation}
{\cal L}_2^{(W)} = \left[ {1 - {{\left( {1 + \tau \frac{{3{{\cal T}_0}}}{{2{T_0}}}} \right)}^{4/9}}} \right]\mathop {\lim }\limits_{\Omega _0^{} \to 1/\ell } \frac{{{{\left( {2\pi \ell {T_0}} \right)}^{2/3}}\sqrt {1 + 3\Omega _0^2{\ell ^2}} }}{{2{{(1 - \Omega _0^2{\ell ^2})}^{5/6}}}}=\infty.
\end{equation}

By comparing the thermodynamic lengths in the considered cases above one can determine in which approach the system has greater probability of fluctuating from one macro state to another with minimal energy loss. In other words, which thermodynamic manifold leads to optimal implementation of quasi-static protocols in ($T,\Omega$) space. The ratios of the computed thermodynamic lengths,
\begin{equation}\label{eqTDLcompar}
\frac{{{{\tilde {\cal L}}^{(G)}}}}{{{{\cal L}^{(G)}}}} = \frac{1}{{\sqrt {{T_0}} }},\qquad \frac{{{{\tilde {\cal L}}^{(G)}}}}{{{{\cal L}^{(W)}}}} = \sqrt {\frac{3}{{{T_0}}}} ,\qquad \frac{{{{\cal L}^{(G)}}}}{{{{\cal L}^{(W)}}}} = \sqrt 3 ,
\end{equation}
compared at $\Omega_0=0$, are independent of the initial rate of temperature change $\mathcal{T}_0$. One notices that for small initial temperatures, $T_0 \ll 1$, the length $\tilde {\cal L}^{(G)}$ in the fixed-$\Omega$ ensemble is greater than both Hessian-Gibbs (grand-canonical) $ {\cal L}^{(G)}$ and Weinhold's length $ {\cal L}^{(W)}$. In this case, it is less probable for the system to fluctuate into neighboring states, if it is realized in the fixed-$\Omega$ ensemble. For large temperatures, $T_0 \gg 1$, $\tilde {\cal L}^{(G)}$ is smaller and it becomes more efficient to fluctuate into neighboring states. Finally, Weinhold's length is always smaller than the Hessian-Gibbs length along the chosen geodesics, thus it is more efficient for the implementation of quasi-static protocols. 

\section{Conclusion}
\label{sec 6}

Investigating the thermodynamic properties of various black hole solutions in three dimensions plays an important role in revealing
hidden relations between classical gravitational theories and quantum field theories in general. In this context, motivated by the remarkable dualities between gravitational (string) and gauge field theories \cite{Maldacena:1997re}, also known as the holographic principle, we study the thermodynamic properties of the stationary Lifshitz black hole solution of New Massive Gravity obtained in \cite{Sarioglu:2018rhl}. 

Our findings uncover the suitable Riemannian metrics on the space of equilibrium states of the black hole solution, together with several criteria for thermodynamic stability of the system. Our investigation has been conducted mostly within the framework of Thermodynamic Information Geometry, which takes advantage of differential geometry to
study statistical features of various models.

The first set of restrictions (\ref{eqMetricExistenceCond}) on the parameter space of the stationary Lifshitz black hole comes from its metric (\ref{metric}). The scalar curvature (\ref{eqRicciPhysx}) and other higher-order invariants (\ref{eqKrechInv})--(\ref{eqKarhInv}) are regular everywhere except at $\ell=0$. The curve $\Omega \ell= 1$ is also a regular one with respect to the curvature invariants, suggesting it is only a coordinate singularity in the metric. This is also true in the thermodynamic case, where all corresponding thermodynamic scalar curvatures vanish on this curve. 

The second set of restrictions (\ref{eqLTDS}) comes from imposing local thermodynamic stability on the black hole. However, for values of the angular velocity in the range $1/(\sqrt 3 \ell)\leq \Omega<\ell$, the black hole is locally stable only with respect to $C_\Omega$, but not $C_L$. On the other hand, imposing global stability in the grand-canonical ensemble, we find the same restriction as in (\ref{eqMetricExistenceCond}), while there is a different condition (\ref{eqFcond}) on $\Omega$ in the canonical ensemble. Nevertheless, the restrictions in both ensembles include partially or entirely the condition (\ref{eqLTDS}) for local thermodynamic stability.

Further conditions comes from the admissible thermodynamic metrics on the ($T,\Omega$) equilibrium state space of the black hole solution. Here, several approaches were considered. In Weinhold's case, one requires positive definite metric, which leads to condition (\ref{eqAngVelWeinhold}), where the upper bound values of the angular velocity $\Omega$ are defined by the positive real roots of the cubic equation (\ref{eqCubicOmega}). When one considers the Hessian of the Gibbs free energy one finds the condition (\ref{eqSylvCritGibbs}), which stays within the local and global thermodynamic stability of the black hole. By defining a positive definite metric within the New Thermodynamic Geometry we found that the metric is in one to one correspondence with the covariance matrix from quantum thermodynamics. In our specific case, when utilizing the Gibbs free energy as conjugate thermodynamic potential, the correspondence is valid in the fixed-$\Omega$ ensemble. The resulting condition on the parameter space is the same as in (\ref{eqSylvCritGibbs}). For clarity, we briefly state the results from all cases in Table \ref{table:1}.

\begin{table}[h!]
	\centering
	\begin{tabular}{ | l | c | c | c| c |}
		\hline
		\bf{Cases} & \bf{Conditions} &\bf{ Positive} definite & \bf{Stat. geometry} \\ \hline
		LTDS & $0\leq \Omega< {{\sqrt 3 }}/{{(3\ell) }}$ & heat capacity $C>0$ & -- \\ \hline
		GTDS (Helmholtz) & $\Omega  < {{\sqrt {21}/ }}{{(7\ell) }}$ & concave ($\partial^2 F/{\partial T^2}<0$) & -- \\ \hline
		GTDS (Gibbs) & $0\leq \Omega<{1}/{\ell}$ & concave ($\partial^2 G/{\partial T^2}<0$)& --\\ \hline
		Ruppeiner (${\rm{Hess}}(S)$) & -- & no & -- \\ \hline
		Weinhold (${\rm{Hess}}(E)$) & 0 $\leq \Omega  < \sqrt r_{+}$ & yes & Elliptic ($R^{(W)}>0$) \\
		\hline
		Gibbs (${\rm{Hess}}(G)$) & $0\leq\Omega  < {{\sqrt 3 }}/{{(3\ell) }}$ &  yes & Elliptic ($R^{(G)}>0$)\\ \hline
		Helmholtz (${\rm{Hess}}(F)$) & -- & no & -- \\ \hline
		NTG(Gibbs) & $0\leq\Omega  < {{\sqrt 3 }}/{{(3\ell) }}$ & yes & Flat ($\tilde R^{(G)}=0$) \\ \hline
	\end{tabular}
	\caption{Conditions for thermodynamic stability in different cases. Notations in the table are as follow: LTDS (local thermodynamic stability), GTDS (global thermodynamic stability), NTG (New Thermodynamic Geometry), ${\rm{Hess}}(S)$ hessian of the entropy, $\sqrt r_{+}\approx 0.185/\ell$. In all cases we assume $\ell>0,\,\,T>0$.}
	\label{table:1}
\end{table}

By considering geodesics on the equilibrium manifold one can find the most optimized implementation of quasi-static protocols. This can be achieved by investigating the thermodynamic length along a chosen geodesic path, which we did in Section \ref{sec 5}. Considering constant angular velocity geodesics we found that Weinhold's approach is more efficient than the Gibbs metric from (\ref{eqHessGibbsMetric}). On the other hand the efficiency of the NTG approach depends on the initial temperatures of the black hole. A relative comparison of the thermodynamic lengths in the corresponding approaches is given in Eq. (\ref{eqTDLcompar}).

In addition to our analysis, one can go further and consider logarithmic corrections to the entropy  due to small thermal fluctuations around its equilibrium configuration. It was shown that for any thermodynamic system with well-defined first law one can write the corrected form of the entropy in the form \cite{Kaul:2000kf, Carlip:2000nv, Das:2001ic, More:2004hv}
\begin{equation}
\tilde S = S + \alpha \log(S T^2)+\cdots 
\end{equation}
where $S$ is given in Eq. (\ref{eqS}) and $\alpha$ is an unknown coefficient. It is straightforward to compute the corrected specific heats of the stationary Lifshitz black hole:
\begin{equation}
\tilde C_\Omega = T\left(\frac{{\partial \tilde S}}{{\partial T}}\right)_\Omega = C_\Omega + \frac{7 \alpha}{3},
\end{equation}
where $C_\Omega$ is given in Eq. (\ref{eqC}), and 
\begin{equation}\label{eqCLcorr}
{\tilde C_L}=T{\left( {\frac{{\partial \tilde S}}{{\partial T}}} \right)_L} = C_L + \frac{{\alpha (7+11{\Omega ^2}{\ell ^2} )}}{{3+7{\Omega ^2}{\ell ^2} }},
\end{equation}
where $C_L$ is defined in Eq. (\ref{eqCL}). Assuming $\alpha\neq\pm\infty$, the corrected heat capacities show no additional singularities. Imposing local thermodynamic stability, $\tilde C_{\Omega,L}>0$, one finds two cases. The first case is for $\alpha<0$, where one has
\begin{equation}\label{eqLogStability}
T> \frac{{|\alpha {|^3}{{(1 - {\Omega ^2}{\ell ^2})}^2}{{(7 + 11{\Omega ^2}{\ell ^2})}^3}}}{{16{\pi ^4}{\ell ^4}{{\left( {1 - 3{\Omega ^2}{\ell ^2}} \right)}^3}}},\quad 3\Omega \ell  < \sqrt 3 ,\quad \alpha  < 0.
\end{equation}
Here, local thermodynamic stability requires a specific $\alpha$-dependent relation (\ref{eqLogStability}) between the temperature $T$ and the angular velocity $\Omega$. For bigger values of $|\alpha|$ higher temperatures are necessary to maintain locally stable equilibrium. Therefore, Eq. (\ref{eqLogStability}) can be interpreted as a lower positive bound on $T$.
The second possibility is $\alpha>0$, where one finds
\begin{equation}\label{eqLogStability2}
0 < T < \frac{{{\alpha ^3}{{(1 - {\Omega ^2}{\ell ^2})}^2}{{(7 + 11{\Omega ^2}{\ell ^2})}^3}}}{{16{\pi ^4}{\ell ^4}{{\left( {3{\Omega ^2}{\ell ^2} - 1} \right)}^3}}},\quad 3\Omega \ell  > \sqrt 3 ,\quad \alpha  > 0,
\end{equation}
in which case $T$ acquires an upper bound.

\section*{Acknowledgments}

The authors would like to thank Radoslav Rashkov, Hristo Dimov, Stoycho Yazadjiev, Miroslav Radomirov, Petia Nedkova and Galin Gyulchev for the useful discussions on the manuscript. T. V. is grateful to Seyed Ali Hosseini Mansoori for careful reading of the draft. This work was partially supported by the Bulgarian NSF grants \textnumero~DM18/1 and \textnumero~N28/5, and the Sofia University grant \textnumero~80-10-149. T. V. and K. S. gratefully acknowledge the support of the Bulgarian national program ``Young Scientists and Postdoctoral Research Fellows''. 

\appendix
\section{The Smarr relation and quasi-homogeneity of the energy }\label{appA}

One can verify the following Smarr relation between the relevant thermodynamic quantities
\begin{equation}
E=\frac{1}{4}TS+\Omega L.
\end{equation}
It is a consequence of Euler’s theorem for quasi-homogeneous function, where we can simply write down $E = E(S,L)$ as
\begin{equation}
E(S,L) = \frac{1}{4}\frac{{\partial E}}{{\partial S}}S + \frac{{\partial E}}{{\partial L}}L.
\end{equation}
Hence, under re-scaling of the form $S\to\alpha S$ and $L\to\alpha^4 L$ with a parameter $\alpha$, one has $E(\alpha S,\alpha^4 L)=\alpha^4 E(S,L)$, thus the energy $E$ is a quasi-homogeneous function of degree 4. We can also directly check if $E(S,L)$ is a quasi-homogeneous functions by analyzing the roots of the following cubic equation with respect to $E$
\begin{equation}
{E^3} - \frac{1}{4}{\left( {\frac{S}{{2\pi \ell }}} \right)^4}{E^2} - \frac{{9{L^2}}}{{8{\ell ^2}}}E + {L^2}\left( {\frac{{27{\pi ^4}{L^2}}}{{4{S^4}}} + \frac{{{S^4}}}{{64{\pi ^4}{\ell ^6}}}} \right) = 0.
\end{equation}
Taking into account the various ranges of the parameters one notes that the only real root, which gives the correct $T=\partial E/\partial S$ and $\Omega=\partial E/\partial L$, is given by
\begin{equation}
E(S,L) = \frac{{{S^4}}}{{192{\pi ^4}{\ell ^4}}} + \frac{{{S^{28/3}}}}{{192{\pi ^4}{\ell ^4}\sqrt[3]{{A(S,L)}}}} + \frac{{72{\pi ^4}{\ell ^2}{L^2}{S^{4/3}}}}{{\sqrt[3]{{A(S,L)}}}} + \frac{{\sqrt[3]{{A(S,L)}}}}{{192{\pi ^4}{\ell ^4}{S^{4/3}}}},
\end{equation}
where the auxiliary function $A(S,L)$ is defined by
\begin{align}
\nonumber A(S,L) &= 192{\pi ^4}{\ell ^3}L{S^8}\left( {\sqrt 3 \sqrt {1728{\pi ^8}{\ell ^6}{L^2} - {S^8}}  - 180{\pi ^4}L{\ell ^3}} \right)\\
&- 331776{\pi ^{12}}{\ell ^9}{L^3}\left( {\sqrt 3 \sqrt {1728{\pi ^8}{\ell ^6}{L^2} - {S^8}}  + 72{\pi ^4}{\ell ^3}L} \right) + {S^{16}},
\end{align}
Let us re-scale the arguments $S\to\alpha S$ and $L\to\alpha^4 L$, hence the energy becomes 
\begin{equation}\label{eqEnergyRescaledAppendix}
E(\alpha S,{\alpha ^4}L) = \frac{{{\alpha ^4}{S^4}}}{{192{\pi ^4}{\ell ^4}}} + \frac{{{\alpha ^{28/3}}{S^{28/3}}}}{{192{\pi ^4}{\ell ^4}\sqrt[3]{{A(\alpha S,{\alpha ^4}L)}}}} + \frac{{72{\pi ^4}{\ell ^2}{\alpha ^{28/3}}{L^2}{S^{4/3}}}}{{\sqrt[3]{{A(\alpha S,{\alpha ^4}L)}}}} + \frac{{\sqrt[3]{{A(\alpha S,{\alpha ^4}L)}}}}{{192{\pi ^4}{\ell ^4}{\alpha ^{4/3}}{S^{4/3}}}}.
\end{equation}
The new function $A(\alpha S,{\alpha ^4}L)$ yields
\begin{align*}
A(\alpha S,{\alpha ^4}L) &= 192{\pi ^4}{\ell ^3}{\alpha ^{12}}L{S^8}\left( {\sqrt 3 \sqrt {1728{\pi ^8}{\ell ^6}{\alpha ^8}{L^2} - {\alpha ^8}{S^8}}  - 180{\pi ^4}{\alpha ^4}L{\ell ^3}} \right)\\
&- 331776{\pi ^{12}}{\ell ^9}{\alpha ^{12}}{L^3}\left( {\sqrt 3 \sqrt {1728{\pi ^8}{\ell ^6}{\alpha ^8}{L^2} - {\alpha ^8}{S^8}}  + 72{\pi ^4}{\ell ^3}{\alpha ^4}L} \right) + {\alpha ^{16}}{S^{16}}\\
&= {\alpha ^{16}}\left[ {192{\pi ^4}{\ell ^3}L{S^8}\left( {\sqrt 3 \sqrt {1728{\pi ^8}{\ell ^6}{L^2} - {S^8}}  - 180{\pi ^4}L{\ell ^3}} \right)} \right.\\
&\left. { - 331776{\pi ^{12}}{\ell ^9}{L^3}\left( {\sqrt 3 \sqrt {1728{\pi ^8}{\ell ^6}{L^2} - {S^8}}  + 72{\pi ^4}{\ell ^3}L} \right) + {S^{16}}} \right]\\
&= {\alpha ^{16}}A(S,L).
\end{align*}
Substituting the last expression in Eq. (\ref{eqEnergyRescaledAppendix}) one finds
\begin{align*}
E(\alpha S,{\alpha ^4}L) &= \frac{{{\alpha ^4}{S^4}}}{{192{\pi ^4}{\ell ^4}}} + \frac{{{\alpha ^{28/3}}{S^{28/3}}}}{{192{\pi ^4}{\ell ^4}{\alpha ^{16/3}}\sqrt[3]{{A(S,L)}}}} + \frac{{72{\pi ^4}{\ell ^2}{\alpha ^{28/3}}{L^2}{S^{4/3}}}}{{{\alpha ^{16/3}}\sqrt[3]{{A(S,L)}}}} + \frac{{{\alpha ^{16/3}}\sqrt[3]{{A(S,L)}}}}{{192{\pi ^4}{\ell ^4}{\alpha ^{4/3}}{S^{4/3}}}}\\
&= {\alpha ^4}\left( {\frac{{{S^4}}}{{192{\pi ^4}{\ell ^4}}} + \frac{{{S^{28/3}}}}{{192{\pi ^4}{\ell ^4}\sqrt[3]{{A(S,L)}}}} + \frac{{72{\pi ^4}{\ell ^2}{L^2}{S^{4/3}}}}{{\sqrt[3]{{A(S,L)}}}} + \frac{{\sqrt[3]{{A(S,L)}}}}{{192{\pi ^4}{\ell ^4}{S^{4/3}}}}} \right)\\
&= {\alpha ^4}E(S,L).
\end{align*}
Therefore, the energy is a quasi-homogeneous function of degree 4, $E(\alpha S,\alpha^4 L)=\alpha^4 E(S,L)$.





\bibliographystyle{utphys}
\bibliography{TS-refs}

\end{document}